\documentclass[aps,pra,twocolumn,a4paper,showpacs,superscriptaddress,10pt]{revtex4-1}

\usepackage{amsmath,amssymb}
\usepackage{ulem,graphicx,xcolor}
\usepackage[colorlinks=true,urlcolor=blue,citecolor=blue,linkcolor=blue]{hyperref}

\begin{document}

\title{Higher-order nonclassical effects in fluctuating-loss channels}

\author{M. Bohmann}\email{martin.bohmann@uni-rostock.de}
\affiliation{Institut f\"ur Physik, Universit\"at Rostock, Albert-Einstein-Stra\ss e 23, D-18051 Rostock, Germany}

\author{J. Sperling}
\affiliation{Clarendon Laboratory, University of Oxford, Parks Road, Oxford OX1 3PU, United Kingdom}

\author{A. A. Semenov}
\affiliation{Institut f\"ur Physik, Universit\"at Rostock, Albert-Einstein-Str. 23, D-18051 Rostock, Germany}
% \affiliation{Institute of  Physics, NAS of Ukraine, Prospect Nauky 46, UA-03028 Kiev, Ukraine}

\author{W. Vogel}
\affiliation{Institut f\"ur Physik, Universit\"at Rostock, Albert-Einstein-Str. 23, D-18051 Rostock, Germany}

\begin{abstract}
	We study the evolution of higher-order nonclassicality and entanglement criteria in atmospheric fluctuating-loss channels.
	By formulating input-output relations for the matrix of moments, we investigate the influence of such channels on the corresponding quantumness criteria.
	This generalization of our previous work on Gaussian entanglement [Bohmann {\sl et al.}, Phys. Rev. A {\bf 94}, 010302(R) (2016)] not only exploits second-order-based scenarios, but it also provides a detailed investigation of  nonclassicality and entanglement in non-Gaussian and multimode radiation fields undergoing a fluctuating attenuation.
	That is, various examples of criteria and states are studied in detail, unexpected effects, e.g., the dependency of the squeezing transfer on the coherent displacement, are discovered,
	and it is demonstrated that non-Gaussian entanglement can be more robust against atmospheric losses than Gaussian one.
	Additionally, we propose a detection scheme for measuring the considered moments after propagation through the atmosphere.
	Therefore, our results may help to develop, improve, and optimize non-Gaussian sources of quantum light for applications in free-space quantum communication.	
\end{abstract}

\date{\today}

\maketitle

\section{Introduction}
	Today, the insecure exchange of confidential or personal data is one of the major challenges to be overcome.
	For this very reason, secure communication based on quantum key distribution, quantum communication protocols for sharing information over large distances, and novel sources of quantum correlated light have attracted our attention over the past years.
	Beside the traditional communication link, employing optical fibers and repeaters, and due to recent experimental advances, free-space channels have turned out to be a promising contender for the transmission of quantum light.
	Implementations of such atmospheric quantum channels were first realized for horizontal ground-to-ground communication~\cite{Ursin, Scheidl, Fedrizzi2009, Capraro, Yin, Ma, Peuntinger}.
	Yet, recent experiments have also promoted the field of atmospheric ground-to-satellite quantum optics~\cite{Villoresi08,Vallone15,Dequal16,Vallone16}.

	In order to implement and optimize quantum communication in atmospheric links, it is crucial to study the influence of fluctuating losses on the quantum properties of radiation fields.
	Such an analysis requires the understanding of two principal aspects:
	first, how atmospheric loss mechanisms work and how they alter the quantum state of light, and second, how the quantum features of light vary under such disturbances and under which conditions quantumness might be preserved.

	On the one hand, a consistent quantum theory of atmospheric losses has been introduced in Ref.~\cite{Semenov2009}.
	Based on this theory, different fluctuating-loss models have been introduced, for example, for weak turbulence, dominated by the effect of beam wandering~\cite{beamwandering}, as well as for weak-to-moderate and strong turbulence, governed by beam-shape deformation and beam-broadening~\cite{VSV2016}.
	Those theoretical descriptions are in good agreement with actual experimental data~\cite{Usenko,VSV2016}.
	Additionally, the techniques of balanced homodyne~\cite{Elser,Semenov2012} and heterodyne~\cite{Croal16} detection have been adapted for turbulent atmospheric channels by propagating the signal and the local oscillator in orthogonal polarization modes.
	This enables one to measure quantum light in such systems even in the presence of bright day light.

	On the other hand, quantum effects after passing light through atmospheric channels have been profoundly examined---but only for a restricted class of scenarios.
	This includes successful studies of the violation of Bell inequalities~\cite{Ursin,Semenov2010}, quantum teleportation~\cite{Yin,Ma}, squeezed light~\cite{Peuntinger}, and entanglement of Gaussian~\cite{GaussSatellites} and non-Gaussian states~\cite{Bohmann15}.
	Still, much more research has to be done in order to get a deeper and more universal understanding of quantum states that are subjected to atmospheric fluctuations.
	We have taken one step in this direction with our recent treatment of bipartite Gaussian entangled states for arbitrary fluctuating-loss channels~\cite{BSSV16}.
	In particular, we introduced an input-output relation for the covariance matrix and the corresponding Simon entanglement test~\cite{Simon2000}, which allowed us to determine for which states and under which atmospheric conditions Gaussian entanglement can be preserved.
	
	For the identification of nonclassicality of radiation fields various criteria in terms of different operator moments have been introduced.
	This includes criteria based on photon number moments~\cite{AgTa}, nonclassicality inequalities in terms of intensity moments based on majorization~\cite{Majorization}, and quadrature moments~\cite{Ag}.
	Furthermore, complete tests based on the negativity of the matrix of moments have been formulated~\cite{SV05nca,SV05ncb,Miran}.
	For the case of nonclassicality, the negativity of the normally ordered matrix of moments indicates the quantum character of the corresponding state~\cite{SV05nca,SV05ncb}.
	In a similar way one can identify bi- and multipartite entanglement in terms of the negativity of the partial transposition (NPT) through the partially transposed matrix of moments~\cite{ShchukinVogel2005,SV06RC}.
	
	In this article, we perform a rigorous analysis of different quantum effects in the presence of a fluctuating-loss medium.
	This significantly extends our previous study to single and multimode nonclassicality as well as non-Gaussian and multimode entanglement.
	For this purpose, we will employ criteria based on the matrices of moments. 
	By deriving input-output relations for those methods, we obtain rather general criteria for nonclassicality and NPT entanglement under atmospheric conditions. 
	Thereby, we can formulate hierarchies of nonclassicality and entanglement conditions based on the minors of the resulting matrices of moments.
	Furthermore, we introduce fluctuation parameters that quantify the influence of fluctuating losses. 
	Depending on those parameters, we investigate the survival of quantum effects in atmospheric links and explicitly derive  conditions for the preservation of different quantum features.
	Furthermore, we will propose a measurement strategy to describe how those desired turbulent moments can be inferred.

	The article is structured as follows.
	In Sec.~\ref{ch:moments}, we derive general input-output relations for the single- and multimode matrices of moments for turbulent loss media.
	Furthermore, fluctuation parameters are introduced.
	Single-mode nonclassicality is studied in Sec.~\ref{ch:ncm}.
	In Sec.~\ref{ch:multi-mode-nc}, we extend this treatment to the multipartite case.
	General NPT-entanglement conditions are examined in Sec.~\ref{ch:multimode}.
	In Sec.~\ref{ch:measurement}, a measurement strategy for obtaining the moments after passing atmospheric channels is proposed.
	A summary can be found in Sec.~\ref{ch:Summary}.

\section{Input-output relations for moments}\label{ch:moments}
	In this section, we relate the moments of annihilation and creation operators after propagating in fluctuating-loss channels to the unperturbed ones by formulating general input-output relations for the single- and multimode case.
	In particular, we will formulate such relations for the matrices of moments, which eventually allows us to treat general non-Gaussian quantum correlations of radiation fields.
	At the end of this section, fluctuation parameters are introduced which will be used in the further course of the article.

\subsection{Single-mode matrix of moments}
	We start our analysis for a single mode which is described by the annihilation (creation) operator $\hat a$ $(\hat a^\dagger)$ of the considered mode.
	Normally ordered moments of these operators transform as~\cite{Semenov2009}
	 \begin{equation}\label{eq:momentsinout}
		\langle \hat{a}^{\dagger n}\hat{a}^{m}\rangle_\mathrm{out}
		=\langle T^{n+m}\rangle\langle \hat{a}^{\dagger n}\hat{a}^{m}\rangle,
        \end{equation}
	where the expectation value $\langle\, \cdot\,\rangle_{\rm out}$ denotes the moments at the output of the fluctuating-loss channel and $T$ is the real-valued amplitude transmission coefficient of the fluctuating-loss channel.
	The moments $\langle T^{l}\rangle$, with $l\in\mathbb{N}$, are given by
	\begin{align}
		\langle T^{l}\rangle=\int_0^1 dT\,\mathcal{ P}(T) T^{l},
	\end{align}
	where $\mathcal{ P}(T)$ is the probability distribution of the transmission coefficient~\cite{Semenov2009}, which accounts for the effects of turbulence in the propagation media and a finite receiver aperture.
	It has been shown that various atmospheric channels can be described by an appropriate $\mathcal{P}(T)$~\cite{beamwandering,VSV2016}.
	Note that the deterministic loss case with a fixed transmission coefficient $T_0$ corresponds to $\mathcal{ P}(T)=\delta(T-T_0)$, where $\delta$ is the Dirac $\delta$ distribution.
	Let us consider a single-mode operator function 
	\begin{align}
		\hat f=\sum_{p,q=0}^\infty f_{p,q}\hat a^{\dagger p} \hat a^q.
	\end{align}
	With the use of this function, we can define the matrix of moments $M_{(p,q),(r,s)}^{\mathrm{out}}$, which includes the fluctuating-loss effects, through
	\begin{align}\label{eq:ff}
		\langle \hat f^\dagger \hat f\rangle_{\mathrm{out}}=\sum_{p,q,r,s}f_{p,q}^*f_{r,s}\underbrace{\langle [\hat a^{\dagger p} \hat a^q]^\dagger[\hat a^{\dagger r} \hat a^s]\rangle_{\mathrm{out}}}_{=M_{(p,q),(r,s)}^{\mathrm{out}}}.
	\end{align}
	The unperturbed matrix of moments at the input is denoted as $M_{(p,q),(r,s)}=\langle [\hat a^{\dagger p} \hat a^q]^\dagger[\hat a^{\dagger r} \hat a^s]\rangle$.

	In order to determine the elements of the matrix of moments $M_{(p,q),(r,s)}^{\mathrm{out}}$ at the output of the turbulent channel, we have to determine how they scale with the moments of the fluctuating loss.
	Therefore, we bring them in the normally ordered form first, which allows us to apply Eq.~\eqref{eq:momentsinout}, and then we infer the scaling with the moments of the transmission coefficient.
	To obtain the input-output relation, we reorganize the moments back into their original ordering.
	For doing so, we use the transformations
	\begin{align}\nonumber
		\hat a^m \hat a^{\dagger n}&=\sum_{i=0}^{\min(m,n)}\frac{m!n!}{i!(m-i)!(n-i)!}\hat a^{\dagger n-i}\hat a^{m-i}\quad\text{and}\\
		\hat a^{\dagger m} \hat a^{ n}&=\sum_{i=0}^{\min(m,n)}\frac{m!n!(-1)^i}{i!(m-i)!(n-i)!}\hat a^{n-i}\hat a^{\dagger m-i}.
		\label{eq:ordering}
	\end{align}

	Applying those steps, we finally obtain the input-output relation
	\begin{align}\label{eq:Mout}
	\begin{aligned}
		M_{(p,q),(r,s)}^{\mathrm{out}}=&\sum_{k=0}^{\min(p,r)}
		\frac{p!r!}{k!(p-k)!(r-k)!}
		\\&\times \left\langle T^{q+r+s+p-2k}(1-T^2)^{k} \right\rangle
		\\&\times M_{(p-k,q),(r-k,s)}.
	\end{aligned}
	\end{align}
	This represents the general input-output relation for the single-mode matrix of moments.
	It is important to notice that the elements of the output matrix of moments are given by a mixture of lower order input elements which are scaled by different moments of the transmission coefficient, cf. Eq.~\eqref{eq:Mout}.
	Based on this relation, we will be able to formulate single-mode noncassicality tests under general atmospheric turbulence conditions.
	
\subsection{Multimode matrix of moments}
	Now we will formulate input-output relations for multipartite cases.
	Therefore, we consider a multimode operator function
	\begin{align}\label{eq:f_multi}
		\hat f=\sum_{\vec p,\vec q=0}^\infty f_{\vec p, \vec q}\hat{ \vec{ a}}^{\dagger \vec{p}} \hat{\vec{a}}^{\vec{q}}.
	\end{align}
	Here we use the $N$-dimensional multi-index notation.
	That includes the definitions $\vec x^{\vec k}=x_1^{k_1}\times\cdots\times x_N^{k_N}$ and $\vec k!=k_1!\times\cdots\times k_N!$.
	Similar to Eq.~\eqref{eq:ff}, we define the multipartite matrix of moments $M_{(\vec p,\vec q),(\vec r,\vec s)}=\langle [\hat{\vec{a}}^{\dagger \vec{p}} \hat{\vec{a}}^{\vec{q}}]^\dagger[\hat{\vec{a}}^{\dagger \vec{r}} \hat{\vec{a}}^{\vec{s}}]\rangle$ and its output version $M_{(\vec p,\vec q),(\vec r,\vec s)}^{\rm out}$.
	In a straightforward manner, the transformations~\eqref{eq:ordering} can be extended to the multipartite scenario.
	Then, in its most general form, the output expression for the multipartite matrix of moments reads as
	\begin{align}\label{eq:multiM}
	\begin{aligned}
		M_{(\vec p,\vec q),(\vec r,\vec s)}^{\mathrm{out}}=&\sum_{\vec k=\vec 0}^{\min(\vec p,\vec r)}\frac{\vec p!\vec r!}{\vec k!(\vec p-\vec k)!(\vec r-\vec k)!}
		\\&\times\left\langle \vec T^{\vec q+\vec r+\vec s+\vec p-2\vec k}\left(\vec 1-\vec T^2\right)^{\vec k}\right\rangle 
		\\&\times M_{(\vec p-\vec k,\vec q),(\vec r-\vec k,\vec s)},
	\end{aligned}
	\end{align}
	where the functions $x\mapsto x^2$ and $(x,y)\mapsto \min(x,y)$ act element-wise on vectors and $\vec 1=(1,\dots,1)^\mathrm{T}$.
	The moments of the transmission coefficients are given by
	\begin{align}\label{eq:Tmomentsmulti}
		\langle \vec T^{\vec l}\rangle&=\int_{[0,1]^N} d^N\vec T\,\mathcal{ P}(\vec T)\, \vec T^{\vec l}\\
		&=\int_0^1 dT_1\dots \int_0^1 dT_N\,\mathcal{P}(T_1,\dots,T_N) T^{l_1}\dots T^{l_N}.\nonumber
	\end{align}

	In contrast to the single mode case, differently correlated loss scenarios can occur in the multipartite case.
	The channels might be totally correlated, which means $T_1=\dots=T_N=T$, $\mathcal{P}(\vec T)=\mathcal{P}(T)$, and $\langle \vec T^{\vec l}\rangle=\langle T^{l_1+\dots +l_N}\rangle$.
	An example of such a correlated scenario might be different optical modes co-propagating through the same atmospheric channel within its correlation time~\cite{Semenov2012}.
	The other extreme is a totally uncorrelated channel.
	That is, $\mathcal{P}(\vec T)=\prod_{i=1}^{N}\mathcal{P}_i(T_i)$ and $\langle \vec T^{\vec l}\rangle=\prod_{i=1}^{N}\langle T_i^{l_i}\rangle$.
	This is usually the case when the different optical modes propagate in different spatial directions through the atmosphere.
	As Eq.~\eqref{eq:Tmomentsmulti} does not make any restrictions to the particular turbulence of the atmosphere, it also describes, e.g., partly correlated channels.
	In the following, we will also study those different cases as this may significantly influence the nonclassical properties of the output fields.
	Based on the input-output relation~\eqref{eq:multiM}, we will be able to analyze quantum properties of multipartite radiation fields evolving in fluctuating-loss channels.
	
\subsection{Fluctuation parameters}
	Before we study different nonclassical effects in fluctuating-loss scenarios, we define different correlation parameters $\Gamma$ for the considered loss scenarios.
	Such quantities will serve as a measure for the turbulence strength, especially, when one uses criteria based on $2\times 2$ minors of the matrix of moments.
	In the single mode case, the fluctuation parameter $\Gamma$ can be defined as
	\begin{align}\label{eq:Gamma_singlemode}
		\Gamma^{(k)}=\frac{\langle T^{2k}\rangle-\langle T^{k}\rangle^2}{\langle T^{2k}\rangle}=\frac{\langle(\Delta T^k)^2\rangle}{\langle T^{2k}\rangle}=1-\frac{\langle T^{k}\rangle^2}{\langle T^{2k}\rangle},
	\end{align}
	with $k\in\mathbb{N}\setminus \{0\}$.
	$\Gamma^{(k)}$ is the ratio between the $k$-th order variance $\langle(\Delta T^k)^2\rangle$ and the moment $\langle T^{2k}\rangle$.
	Hence, it is a measure for the influence of the turbulence, i.e., how much $\langle T^{k}\rangle^2$ and $\langle T^{2k}\rangle$ differ from each other.
	It is easy to see that $\Gamma^{(k)}\in[0,1]$.
	In the absence of turbulence, $\langle(\Delta T^k)^2\rangle=0$, $\Gamma^{(k)}$ is zero.
	Therefore, the fluctuation effects are related to a nonzero value of $\Gamma^{(k)}$.
	We have that as the distribution $\mathcal{P}(T)$ is broader, the fluctuating-loss effects are higher---i.e., $\Gamma^{(k)}$ increases, which may also lead to a stronger impact on the nonclassicality of light.
	
	As we will also discuss multimode matrices of moments and their application in Secs.~\ref{ch:multi-mode-nc} and~\ref{ch:multimode}, we will generalize the fluctuation parameters.
	Therefore, we will again use the multi-index notation as introduced above.
	Moreover, a multi-index $\vec k$ may be split into two parts, $A$ and $B$, such that its elements are disjointly distributed in $\vec k_A$ and $\vec k_B$.
	We define
	\begin{align}\label{eq:Gammamulti}
		\Gamma^{(k)}_{\{A;B\}}=\frac{\langle \vec T_A^{2\vec k_A}\rangle\langle \vec T_B^{2\vec k_B}\rangle-\langle \vec T^{\vec k}\rangle^2}{\langle \vec T_A^{2\vec k_A}\rangle\langle \vec T_B^{2\vec k_B}\rangle}.
	\end{align}
	The subscript indicates the two partitions $A$ and $B$ into which the $2\vec k$-th moments can be separated.
	For example, let us consider a bipartite case where the modes are labeled by 1 and 2.
	In this case, there are two possible correlation parameters:
	$\Gamma^{(\vec k)}_{\{1;2\}}=(\langle T_1^{2 k}\rangle\langle T_2^{2 k}\rangle-\langle \vec T^{\vec k}\rangle^2)/(\langle T_1^{2 k}\rangle\langle T_2^{2 k}\rangle)$ for $A=\{1\}$ and $B=\{2\}$ or $\Gamma^{(\vec k)}_{\{1,2\}}=(\langle \vec T^{2\vec k}\rangle-\langle \vec T^{\vec k}\rangle^2)/\langle \vec T^{2\vec k}\rangle$ for $A=\{1,2\}$ and $B=\{\}$.
	The multimode $\Gamma$ differs form the single-mode one in this additional feature.

\section{Single-mode Nonclassical moments}\label{ch:ncm}
	Let us now study nonclassical effects in fluctuating-loss channels for a single mode.
	Therefore, we consider the normally ordered expectation value of the operator function in Eq.~\eqref{eq:ff}.
	It can be given in terms of the Glauber-Sudarshan $P$ function~\cite{Glauber,Sudarshan}
	\begin{align}\label{eq:ffnor}
		\langle:\hat{f}^\dagger \hat{f}:\rangle = \int |f(\alpha)|^2 P(\alpha)\,d^2\alpha.
	\end{align}
	If the $P$ function fails to be interpreted as a probability distribution, the corresponding quantum state is considered to be nonclassical in the sense that it cannot be interpreted as a classical statistical mixture of coherent states. 
	This implies that the expectation value~\eqref{eq:ffnor} is non-negative for any classical state.
	Hence, $\langle{:}\hat{f}^\dagger \hat{f}{:}\rangle<0$ directly indicates the nonclassicality of the considered quantum state~\cite{SV05nca}.
	The expectation value~\eqref{eq:ffnor} can be related to the matrix of moments
	\begin{align}\label{eq:singlemode_nonclassicality}
		\langle:\hat{f}^\dagger \hat{f}:\rangle = \sum_{p,q,r,s}f_{p,q}^*f_{r,s}N_{(p,q),(r,s)},
	\end{align}
	with 
	\begin{align}
		N_{(p,q),(r,s)}=\langle \hat a^{\dagger p+r}\hat a^{q+s}\rangle.
	\end{align}
	
	Note that we use the notation $N_{(p,q),(r,s)}$ for the normally ordered matrix of moments in order to distinguish it from the non-normally ordered matrix of moments~\eqref{eq:ff}.
	In Ref.~\cite{SV05ncb}, it has been shown that a quantum state is nonclassical if and only if its normally ordered matrix of moments is not positive semidefinite.
	In particular, this implies that if one of the principle minors of $N_{(p,q),(r,s)}$ is negative, the nonclassical character of the corresponding radiation field is verified.
	The output matrix of moments is given by
	\begin{widetext}
	\begin{align}\label{eq:single_mode_M_tur}
		N_{(p,q),(r,s)}^{\rm out} =
		\begin{pmatrix}
		   1 & \langle T \rangle \langle \hat{a} \rangle & \langle T \rangle \langle \hat{a}^\dagger \rangle & \langle T^2 \rangle \langle \hat{a}^2 \rangle & \langle T^2 \rangle \langle \hat{a}^\dagger\hat{a} \rangle & \ldots \\
		  \langle T \rangle \langle \hat{a}^\dagger \rangle & \langle T^2 \rangle \langle \hat{a}^\dagger\hat{a}  \rangle &
		  \langle T^2 \rangle \langle \hat{a}^{\dagger 2} \rangle & \langle T^3 \rangle \langle \hat{a}^\dagger \hat{a}^2 \rangle & \langle T^3 \rangle \langle \hat{a}^{\dagger 2} \hat{a} \rangle & \ldots \\
		  \langle T \rangle \langle \hat{a} \rangle & \langle T^2 \rangle \langle \hat{a}^2 \rangle &
		  \langle T^2 \rangle \langle \hat{a}^\dagger\hat{a} \rangle & \langle T^3 \rangle \langle \hat{a}^3 \rangle & \langle T^3 \rangle \langle \hat{a}^\dagger \hat{a}^2 \rangle &  \ldots \\
		  \langle T^2 \rangle \langle \hat{a}^{\dagger 2} \rangle & \langle T^3 \rangle \langle \hat{a}^{\dagger 2} \hat{a} \rangle & \langle T^3 \rangle \langle \hat{a}^{\dagger 3} \rangle & \langle T^4 \rangle \langle \hat{a}^{\dagger 2}\hat{a}^2 \rangle & \langle T^4 \rangle \langle \hat{a}^{\dagger 3} \hat{a} \rangle & \ldots \\
		  \langle T^2 \rangle \langle \hat{a}^\dagger\hat{a} \rangle & \langle T^3 \rangle \langle \hat{a}^\dagger \hat{a}^2 \rangle & \langle T^3 \rangle \langle \hat{a}^{\dagger 2} \hat{a} \rangle & \langle T^4 \rangle \langle \hat{a}^\dagger \hat{a}^3 \rangle & \langle T^4 \rangle \langle \hat{a}^{\dagger 2}\hat{a}^2 \rangle & \ldots \\ 
		  \vdots & \vdots & \vdots & \vdots & \vdots & \ddots
		\end{pmatrix}.
	\end{align}
	\end{widetext}

	Due to the normally ordered form of the nonclassicality condition, each element of $N_{(p,q),(r,s)}$ transforms as given in Eq.~\eqref{eq:momentsinout}, which can be directly applied without any reordering of the moments.
	It is important to observe that the scaling of the different elements of $N_{(p,q),(r,s)}^{\rm out}$ by different moments of the transmission coefficient leads to a distinct and nontrivial dependency on the fluctuation parameters.
	However, due to the normally ordered form [cf. Eq~\eqref{eq:singlemode_nonclassicality}], the output is not a mixture of the input elements, as it occurs in the general (non-normally-ordered) case Eq.~\eqref{eq:Mout}.
	Therefore, it is of great interest to study under which conditions the corresponding nonclassicality test is still sensitive to quantum features.
	Note that any constant loss, except for full loss $(T=0)$, will always preserve the nonclassical features of any quantum state.
	In the following, we will examine the influence of fluctuating loss on two particular nonclassicality condition.

\subsection{Sub-Poissonian light}
	We first examine the notion of sub-Poissionian light, for the first experimental verification, see Ref.~\cite{Short}.
	This form of nonclassicality is based on the photon number distribution of the considered light field.
	Note that sub-Poissionian light in atmospheric channels has also been studied in Ref.~\cite{Semenov2009}.
	If a quantum state shows a photon number distribution which is narrower than a Poissionian distribution, i.e., in which the photon number variance is less than the mean photon number, it is called sub-Poissonian.
	This effect can be identified by the so-called Mandel $Q_{\rm M}$ parameter~\cite{MaQ},
	\begin{align}\label{eq:QM}
		Q_{\rm M}=\frac{\langle(\Delta\hat n)^2\rangle-\langle\hat n\rangle}{\langle\hat n\rangle} \stackrel{\text{sub-Poissonian}}{<} 0.
	\end{align}
	That is, the sub-Poissionian character is then indicated by $Q_{\rm M}$ being negative, which implies that the numerator in Eq.~\eqref{eq:QM} has to be negative. 
	This can be equivalently expressed by a negative subdeterminant,
	\begin{align}\label{eq:Poisson}
		d=
			\begin{vmatrix} 
			1 & \langle\hat a^\dagger\hat a\rangle\\
			\langle\hat a^\dagger\hat a\rangle & \langle\hat a^{\dagger 2}\hat a^2\rangle
			\end{vmatrix}
		<0,
	\end{align}
	including bosonic operators up to the fourth order.

	Let us now consider the output condition after passing through a fluctuating-loss channel.
	It can be formulated via the corresponding minor of the output matrix~\eqref{eq:single_mode_M_tur}
	\begin{align}\label{eq:Poissonout}
		d^{\rm out}=\langle T^4\rangle\left[ d+\Gamma^{(2)}\langle \hat n\rangle^2\right],
	\end{align}
	with $\Gamma^{(2)}$ as defined in Eq.~\eqref{eq:Gamma_singlemode}.
	Let us recall that for $\Gamma^{(2)}=0$ there are no fluctuations of losses, and as $\Gamma^{(2)}$ is larger, the influence of these fluctuations is stronger.
	From Eq.~\eqref{eq:Poissonout} we directly see that an increase of the mean photon number or $\Gamma^{(2)}$ may lead to a disappearance of the sub-Poissonian character, which is due to the positive part added to $d^{\rm out}$.
	In other words, the output nonclassicality condition from the minor~\eqref{eq:Poissonout} is $d<-\Gamma^{(2)}\langle\hat n\rangle^2$, which is more restrictive than the input bound $d<0$.
	Hence, high mean photon numbers are hindering the preservation of sub-Poissonian light in atmospheric channels.

	Let us consider the example of $n$-photon Fock states $|n\rangle$.
	For this case, we readily get that $d=-n$, with $n=\langle \hat n\rangle$, and the output version reads as
	\begin{align}
		d^{\rm out}=\langle T^4\rangle n\left[\Gamma^{(2)}n-1\right].
	\end{align}
	This yields the following direct relation between photon number and the fluctuation parameter: 
	\begin{align}
		n<1/\Gamma^{(2)},
	\end{align}
	for exhibiting a sub-Poissonian character after propagation through a fluctuating-loss channel ($\Gamma^{(2)}\neq0$).
	In particular, as the fluctuating losses (the larger $\Gamma^{(2)}$) were stronger, the photon number $n$ can be smaller in order to identify its sub-Poissionian statistics.

\subsection{$k$th-order amplitude squeezing}
	As a second example, let us now study the nonclassical feature of $k$th power amplitude squeezing in turbulent loss scenarios, which is a generalization of the ordinary notion of squeezing~\cite{Hillery87}.
	The influence of atmospheric losses on quadrature squeezing ($k=1$) has been studied in Ref.~\cite{Semenov2012}.
	In Refs.~\cite{SV05HOAS} and~\cite{SV05ncb}, conditions for $k$th-order amplitude squeezing were introduced in terms of a submatrix of the normally ordered matrix of moments.
	In particular, it can be detected by the negativity of the determinant
	\begin{align}\label{eq:HOAS}
	\begin{aligned}
		d_{k}&=
		\begin{vmatrix} 
			1 & \langle \hat  a^{\dagger k}\rangle & \langle\hat a^k\rangle\\
			\langle\hat a^{k}\rangle & \langle \hat  a^{\dagger k}\hat a^k\rangle & \langle\hat a^{2k}\rangle \\
			\langle\hat a^{\dagger k}\rangle & \langle \hat  a^{\dagger 2k}\rangle & \langle\hat a^{\dagger k}\hat a^{k}\rangle
		\end{vmatrix}
		\\&=
		\begin{vmatrix} 
			\langle\Delta \hat  a^{\dagger k}\Delta\hat a^k\rangle & \langle(\Delta\hat a^k)^2\rangle\\
			\langle(\Delta\hat a^{\dagger k})^2\rangle & \langle\Delta \hat  a^{\dagger k}\Delta\hat a^k\rangle 
		\end{vmatrix},
	\end{aligned}
	\end{align}
	where $\Delta\hat x=\hat x-\langle \hat x \rangle$ and the subscript $k$ in $d_{k}$ indicates the order of the amplitude squeezing.
	If $d_{k}$ is negative, $k$-th power amplitude squeezing is revealed.
	Using the output matrix of moments~\eqref{eq:single_mode_M_tur}, we can formulate the output condition $d_{k}^{\rm out}$ as
	\begin{align}\label{eq:HOASout}
		d_{k}^{\rm out}=&\langle T^{2k}\rangle^2\Big[d_{k}
		+\Gamma^{(k)}(\langle \hat a^{\dagger k}\rangle, \langle\hat a^k\rangle)
		A_{k}
		\begin{pmatrix} 
			\langle \hat a^{k}\rangle \\ \langle\hat a^{\dagger k}\rangle
		\end{pmatrix}
		\Big]<0,
		\\ \nonumber
		&\text{with }
		A_{k}=
		\begin{pmatrix} 
			\langle\Delta \hat  a^{\dagger k}\Delta\hat a^k\rangle & \langle(\Delta\hat a^k)^2\rangle\\
			\langle(\Delta\hat a^{\dagger k})^2\rangle & \langle\Delta \hat  a^{\dagger k}\Delta\hat a^k\rangle 
		\end{pmatrix}
	\end{align}
	and $\Gamma^{(k)}$ defined in Eq.~\eqref{eq:Gamma_singlemode}.

	In Eq.~\eqref{eq:HOASout}, we see that in addition to the scaling of $d_{k}$ by $\langle T^{2k}\rangle^2$, which corresponds to the deterministic loss case, an additional term related to $\Gamma^{(k)}$ is added.
	This turbulence term consists of the quadratic form $(\langle \hat a^{\dagger k}\rangle, \langle\hat a^k\rangle)A_k(\langle \hat a^{\dagger k}\rangle, \langle\hat a^k\rangle)^{\rm T}$.
	Note that the determinant of the matrix $A_k$ is itself the matrix in the input condition $d_k<0$ [cf. Eq.~\eqref{eq:HOAS}].
	Hence, $A_k$ is not positive semidefinite if $d_k<0$, which leads to the fact that the turbulence term can contribute positively as well as negatively to the turbulent condition Eq.~\eqref{eq:HOASout}, depending on the vector $(\langle \hat a^{\dagger k}\rangle,\langle \hat a^{k}\rangle)^\mathrm{T}$ being parallel or perpendicular to the eigenvector to the negative eigenvalue of $A_k$.
	In other words, for the transfer of $k$th-order amplitude squeezing one can find regions of $\langle \hat a^{\dagger k}\rangle$ and $\langle \hat a^{k}\rangle$ for which the nonclassicality is very well preserved while for others it will vanish.
	This is an important finding as such an effect does not occur in deterministic loss scenarios.
	Additionally this gives rise to a state optimization for the nonclassicality transfer in atmospheric channels.
	A similar dependency for two-mode Gaussian entanglement in atmospheric channels has been reported in Ref.~\cite{BSSV16}.

\begin{figure}[ht]
		\includegraphics[width=0.7\linewidth]{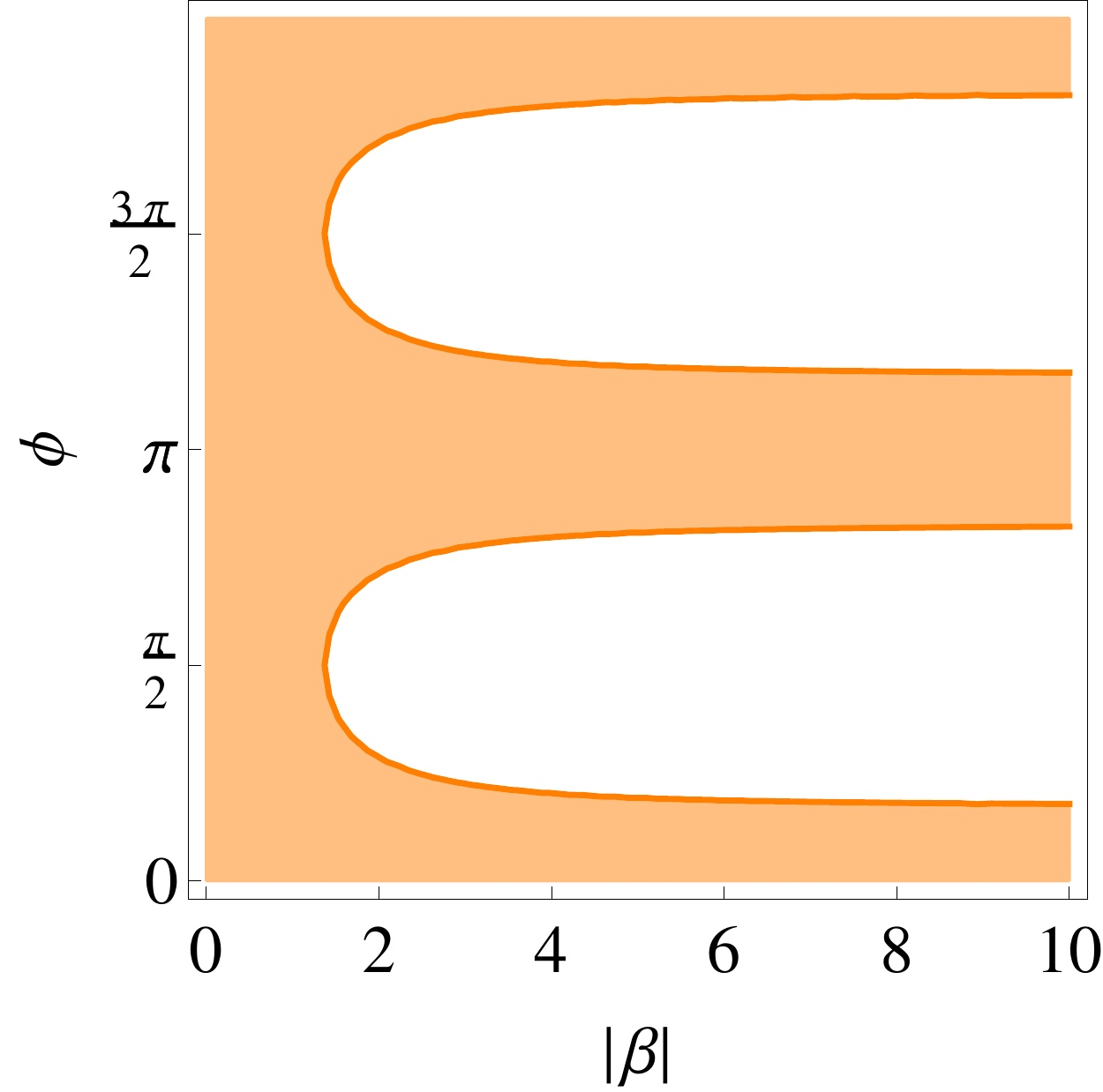}
		\caption{(Color online)
			The test~\eqref{eq:HOASout} ($k=1$) is shown for a displaced and squeezed state $|\xi,\beta\rangle$ depending on the phase $(\phi)$ and amplitude $(|\beta|)$ of its coherent displacement.
			The contours ($d_k^{\rm out}\equiv0$) illustrate the bound of the region for which squeezing can be detected after transmission through a turbulent loss channel.
			Squeezing is preserved ($d_{k}^{\rm out}<0$) in the shaded areas.
			The squeezing parameter is $\xi=0.5$ and the fluctuation parameter is $\Gamma^{(1)}=0.17$.
		}\label{fig:tmsv}
	\end{figure}

	Now we consider the simplest case $k=1$ which is the prominent case of quadrature squeezing, as there occur already interesting phenomena.
	In particular, we study the squeezing transfer through a turbulent channels of a displaced squeezed state $|\xi,\beta\rangle$~\cite{Buch}.
	The state is defined via a real-valued squeezing parameter $\xi$ and coherent displacement $\beta$.
	We analyze the dependency of the nonclassicality test~\eqref{eq:HOASout} for $k=1$ on the phase $\phi=\arg(\beta)$ and absolute value $|\beta|$ of the coherent displacement $\beta=\langle\hat a\rangle=|\beta|e^{i\phi}$, defining the vector of coherent displacement $(\langle \hat a^{\dagger}\rangle, \langle\hat a\rangle)^{\rm T}$.
	Figure~\ref{fig:tmsv} shows that the negativity of the nonclassicality condition~\eqref{eq:HOASout} depends in a nontrivial form on the choice of $|\beta|$ and $\phi$.
	Hence, for the transfer of squeezing through a free-space link, it is possible to find optimal displacement directions in order to preserve the squeezing.
	In particular, zero displacement will always conserve squeezing as then the term proportional to $\Gamma^{(k)}$ in Eq.~\eqref{eq:HOASout} vanishes.

\section{Multimode nonclassical moments}\label{ch:multi-mode-nc}
	Now we will expand the treatment of nonclassical moments including the impact of fluctuating loss to the multipartite case.
	Analogously to the single-mode scenario, the nonclassicality condition reads 
	\begin{align}\label{eq:ffnor_multi}
		\langle:\hat{f}^\dagger \hat{f}:\rangle=\sum_{\vec p,\vec q,\vec r,\vec s}f_{\vec p,\vec q}^*f_{\vec r,\vec s}N_{(\vec p,\vec q),(\vec r,\vec s)}<0,
	\end{align}
	with the multimode operator function $\hat{f}$ defined in Eq.~\eqref{eq:f_multi} and $N_{(\vec p,\vec q),(\vec r,\vec s)}=\langle \hat{\vec{a}}^{\dagger \vec{p}+\vec{r}}\hat{\vec{a}}^{\vec{q}+\vec{s}}\rangle$.
	If $N_{(\vec p,\vec q),(\vec r,\vec s)}$ is negative, nonclassicality of the corresponding multimode system is detected.
	The output normally ordered multimode matrix of moments is given by
	\begin{align}
		N^{\rm out}_{(\vec p,\vec q),(\vec r,\vec s)}=\langle\vec T^{\vec q+\vec r+\vec s+\vec p}\rangle N_{(\vec p,\vec q),(\vec r,\vec s)}.
	\end{align}
	Similar to the single-mode case, different elements of $N_{(\vec p,\vec q),(\vec r,\vec s)}$ are scaled by different moments of transmission coefficients which introduces the fluctuation effects.
	One should bear in mind that in the multipartite case different transmission coefficients are associated to each mode, cf. also Eq.~\eqref{eq:Tmomentsmulti} and the text below it, which results in different scenarios such as correlated and uncorrelated losses.

	Let us study an example of a multipartite nonclassicality test based on the matrix of moments approach with fluctuating losses.
	In particular, we consider a bipartite system via $\langle{:}\hat{f}^\dagger \hat{f}{:}\rangle$ with the choice
	\begin{align}
		\hat f=f_1\hat a^\dagger\hat a+f_2 \hat b^\dagger\hat b.
	\end{align}
	This leads to the nonclassicality condition
	\begin{align}\label{eq:ncorrelation}
		D=
			\begin{vmatrix} 
			\langle \hat a^{\dagger 2}\hat a^2\rangle & \langle\hat a^\dagger\hat a \hat b^\dagger\hat b\rangle\\
			\langle\hat a^\dagger\hat a \hat b^\dagger\hat b\rangle & \langle\hat b^{\dagger 2}\hat b^2\rangle
			\end{vmatrix}<0,
	\end{align}
	which includes fourth-order terms; cf. Ref.~\cite{Agarwal13}.
	Note that inequality~\eqref{eq:ncorrelation} is given by a minor of the general multimode matrix of moments $M_{(\vec p,\vec q),(\vec r,\vec s)}$.
	$D$ indicates if the two considered modes show a joint nonclassical photon number correlation and is structurally similar to the single mode condition~\eqref{eq:Poisson}.

	The output condition---after passing the atmospheric channel---can be given as
	\begin{align}\label{eq:ncorrelationout}
		D_{\rm out}=\langle T_a^4\rangle\langle T_b^4\rangle[D+\Gamma_{\{1;2\}}^{( 2)}\langle\hat a^\dagger\hat a \hat b^\dagger\hat b\rangle^2],
	\end{align}
	with $\Gamma_{\{1;2\}}^{( 2)}$ defined in Eq.~\eqref{eq:Gammamulti}.
	In the case of correlated loss, i.e., $T_a{=}T_b{=}T$ and $\Gamma_{\{1;2\}}^{(2)}=0$, we see that the output test is simply a scaled version of the input condition, $D_{\rm out}=\langle T^4 \rangle^2D$.
	As $D$ indicates a bipartite photon number correlation, it can be understood that in the case of correlated loss, where both modes suffer form the same loss, the initial nonclassicality does not vanish.
	In Ref.~\cite{BSSV16}, we introduced a protocol to artificially correlate uncorrelated turbulent loss channels.
	Applying such an adaptive protocol leads to $\Gamma_{\{1;2\}}^{(2)}=0$, which means that the nonclassical correlation identified by the test~\eqref{eq:ncorrelationout} can be preserved.
	Hence, the protocol can be of practical importance in order to distribute general quantum correlations through atmospheric channels.

	For example, we can consider a fully phase randomized two-mode squeezed-vacuum state
	\begin{align}\label{eq:prtmsv}
		\hat\rho=\sum_{n=0}^\infty(1-p)p^n|n\rangle\langle n|\otimes|n\rangle\langle n|,
	\end{align}
	with $0<p<1$.
	Note that $p$ in this non-Gaussian state is related to the squeezing parameter $\xi$ of the two-mode squeezed-vacuum state by $|\xi|={\rm artanh}(\sqrt{p})=\ln[(1+\sqrt{p})/(1-\sqrt{p})]/2$.
	This state is not entangled, has zero discord, and has a positive Wigner function, and its single-mode reductions are classical.
	However, it is quantum correlated as its $P$ function shows negativities~\cite{Agudelo13}.
	For the state~\eqref{eq:prtmsv}, the condition~\eqref{eq:ncorrelation} yields $D=-p^2(1+p)/(1-p)^3<0$ and its output minor reads 
	\begin{align}\label{eq:ncorrelationturstate}
		D_{\rm out}=\langle T_a^4\rangle\langle T_b^4\rangle\frac{p^2(1+p)}{(1-p)^3}\Big[-1+\Gamma_{\{1;2\}}^{( 2)}\frac{1+p}{1-p}\Big].
	\end{align}
	From this, we can directly identify the condition for revealing quantum correlations by this test, for  $\Gamma_{\{1;2\}}^{( 2)}<(1-p)/(1+p)$.

\begin{figure}[ht]
		\includegraphics[width=0.9\linewidth]{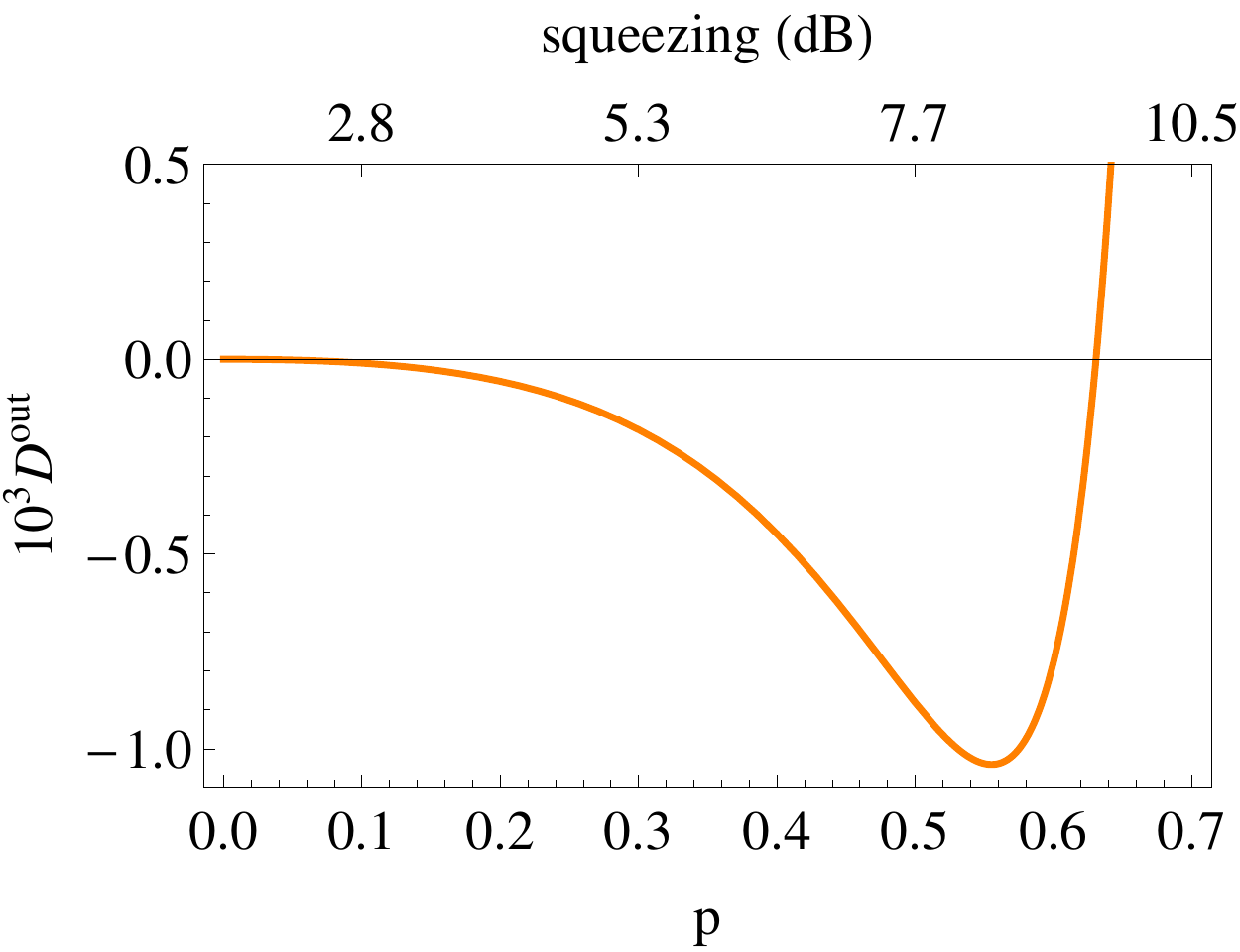}
		\caption{
		(Color online)
		Nonclassicality test~\eqref{eq:ncorrelationturstate} for the non-Gaussian state~\eqref{eq:prtmsv} in an uncorrelated atmospheric fluctuating-loss channel.
		The fluctuation parameters are $\langle T^2_a\rangle=\langle T^2_b\rangle=0.163$, $\langle T^4_a\rangle=\langle T^4_b\rangle=0.030$, and $\Gamma_{\{1;2\}}^{( 2)}=0.23$.
		}\label{fig:prtmsv}
	\end{figure}

	Figure~\ref{fig:prtmsv} shows the nonclassicality test~\eqref{eq:ncorrelationturstate} as a function of the parameter $p$.
	For the fluctuating loss, we use a realistic model which is dominated by the effect of beam wandering~\cite{beamwandering}.
	For a particular choice of parameters in this model (i.e., $a=0.04$\,m, $W=1.5a$, and $\sigma=0.6$; cf. Ref.~\cite{beamwandering}), we get $\langle T\rangle=0.398$, $\langle T^2\rangle=0.163$, and $\langle T^4\rangle=0.030$.
	One can see that the nonclassical correlation can be detected up to $p_{\rm max}=(1-\Gamma_{\{1;2\}}^{( 2)})/(1+\Gamma_{\{1;2\}}^{( 2)})$.
	This can be understood as follows:
	With increasing $p$, the mean photon number of the state~\eqref{eq:prtmsv} increases too.
	Higher photon number contributions, however, are more fragile to fluctuating loss, which leads eventually to the disappearance of the negativity.
	An analog dependency on the mean photon number can be observed in the single-mode case for the condition~\eqref{eq:Poissonout}.
	The entanglement of a two-mode squeezed-vacuum state in fluctuating-loss channels, as studied in Fig.~1 of Ref.~\cite{BSSV16}, shows another similar behavior.
	More precisely, higher squeezing, which corresponds to higher $p$ values in the phase-randomized scenario, is not favorable in atmospheric links and may lead to a classical behavior (see Fig.~\ref{fig:prtmsv}).

\section{Multimode moments and negative partial transposition} \label{ch:multimode}
	In this section, we study entanglement in multimode systems while including fluctuating losses and using the matrix of moments.
	In particular, we will test for entanglement in terms of the Peres-Horodecki criterion~\cite{Peres,Horodecki96}, which is a sufficient condition for entanglement between bipartitions.
	It is verified by the NPT of the corresponding density operator.
	It is worth mentioning that other positive but not completely positive maps could be studied as well.

	Instead of applying the partial transposition to the considered state, one can partially transpose the corresponding subsystems of the matrix of moments in order to obtain moment-based entanglement tests~\cite{ShchukinVogel2005}.
	The negativity of any principle minor of this partially transposed matrix of moments verifies entanglement, yielding an easily accessible hierarchy of entanglement conditions which do not require the reconstruction of the whole density matrix.
	Note that this approach only allows one to test for bipartite entanglement.
	Multipartite entanglement has in general a much richer structure than that given only by bipartitions~\cite{Gerke16}.
	In order to verify such complex entanglement, one may construct suitable multipartite entanglement witnesses~\cite{SV13}, which has been demonstrated for a 10-mode Gaussian frequency comb state~\cite{Gerke15}.

	We start from the multimode matrix of moments $M_{(\vec p,\vec q),(\vec r,\vec s)}=\langle [\hat{\vec{a}}^{\dagger \vec{p}} \hat{\vec{a}}^{\vec{q}}]^\dagger[\hat{\vec{a}}^{\dagger \vec{r}} \hat{\vec{a}}^{\vec{s}}]\rangle$, which is defined by the expectation value $\langle\hat f^\dagger\hat f\rangle$ of the multimode operator function $\hat f$ given in Eq.~\eqref{eq:f_multi}.
	Note that we take here the standard expectation value and not the normally ordered one, as used for the nonclassicality conditions.
	Yet, the entanglement condition can now be formulated by partially transposing the matrix of moments~\cite{ShchukinVogel2005,SV06RC}.
	Therefore, we consider the nonempty subsets $A$ and its complement $B$ of all considered modes in the set $X$, i.e., $A\cap B=\emptyset$ and $A\cup B=X$.
	Applying the transposition with respect to the modes in $B$ yields the partially transposed matrix of moments $M^{\rm PT}_{(\vec p,\vec q),(\vec r,\vec s)}$,
	\begin{align}\label{eq:MPT}
		M^{\rm PT}_{(\vec p,\vec q),(\vec r,\vec s)}=\left\langle\prod_{i\in A}\hat a_i^{\dagger p_i}\hat a_i^{q_i}\hat a_i^{\dagger r_i}\hat a_i^{s_i}\prod_{i\in B}\hat a_i^{\dagger s_i}\hat a_i^{r_i}\hat a_i^{\dagger q_i}\hat a_i^{p_i}\right\rangle.
	\end{align}
	If this matrix is not positive semidefinite, it directly certifies entanglement between the two partitions $A$ and $B$.
	Note that in a multimode scenario, one can test for entanglement between different bipartitions; more precisely, an $N$-mode quantum state can be separated into $2^{N-1}-1$ different nontrivial bipartitions.
	
	Let us now derive the corresponding entanglement test for a state of light that propagates through a fluctuating-loss channel.
	From Eq.~\eqref{eq:multiM}, we obtain the output matrix of moments $M_{(\vec p,\vec q),(\vec r,\vec s)}^{\rm out}$.
	Applying the partially transposition to $M_{(\vec p,\vec q),(\vec r,\vec s)}^{\rm out}$, we get
	\begin{widetext}
	\begin{align}\label{eq:MoutPT}
		M_{(\vec p,\vec q),(\vec r,\vec s)}^{\mathrm{out\,PT}}{=}\sum_{\vec k=\vec 0}^{\min(\vec p,\vec r)}\frac{\vec p!\vec r!}{\vec k!(\vec p-\vec k)!(\vec r-\vec k)!}
		\left\langle \vec T^{\vec q+\vec r+\vec s+\vec p-2\vec k}(\vec 1-\vec T^2)^{\vec k}\right\rangle
		\left\langle\prod_{i\in A}\hat a_i^{\dagger p_i-k_i}\hat a_i^{q_i}\hat a_i^{\dagger r_i-k_i}\hat a_i^{s_i}\prod_{i\in B}\hat a_i^{\dagger s_i}\hat a_i^{r_i-k_i}\hat a_i^{\dagger q_i}\hat a_i^{p_i-k_i}\right\rangle\!.
	\end{align}
	\end{widetext}
	The negativity of principle minors of $M_{(\vec p,\vec q),(\vec r,\vec s)}^{\mathrm{out,\,PT}}$ will reveal entanglement between the two partitions after the fluctuating-loss channel.
	We see that the influence of the fluctuation leads to two effects.
	First, different moments of the matrix of moments are scaled by different moments of the transmission coefficient.
	Second, the sum in Eq.~\eqref{eq:MoutPT} causes a mixing between different elements of the unperturbed matrix of moments.
	Let us emphasize again that Eq.~\eqref{eq:MoutPT} is the most general form of the partially transposed matrix of moments including the effects of fluctuating losses.
	
\subsection{Gaussian entanglement test}
	Let us restrict the partially transposed matrix of moments $M^{\rm PT}$, cf.~Eq.~(\ref{eq:MPT}), to bipartite moments up to the second order. 
	Then, bipartite Gaussian entanglement is identified if and only if
	\begin{align}\label{eq:Simon}
		G=\det{M}^{\rm PT}<0, 
	\end{align}
	which represents the Simon entanglement criterion~\cite{Simon2000} in the form of Ref.~\cite{ShchukinVogel2005}. 

	We gave a complete and rigorous treatment for this Gaussian entanglement in atmospheric channels in Ref.~\cite{BSSV16}. 
	After passing through a fluctuating-loss channel, the structure of entanglement certifier $G$, cf.~Eq.~(\ref{eq:Simon}), can be given in the form
	\begin{align}\label{eq:IOR_Simon}
	G^\mathrm{out}=G_{\langle T_a^2\rangle,\langle T_b^2\rangle}+G^{\rm tur.},
	\end{align}
	where $G^\mathrm{out}$ is the Simon entanglement test for the light at the receivers which can be split into two contributions:
	$G_{\langle T_a^2\rangle,\langle T_b^2\rangle}$ corresponds to the Simon entanglement test for deterministic attenuations with transmission efficiencies $\langle T_a^2\rangle$ and $\langle T_b^2\rangle$, and $G^{\rm tur.}$ is the term which accounts for all effects of fluctuating loss.
	If the initial Gaussian state is entangled, i.e. $G{<}0$ at the transmitter, the term $G_{\langle T_a^2\rangle,\langle T_b^2\rangle}$ is always negative for a quite broad class of entanglement-robust states, cf.~Ref.~\cite{Barbosa}.
	The fluctuation-related term $G^{\rm tur.}$, cf.~Ref.~\cite{BSSV16} for its explicit form, depends on the first and second moments of the transmission coefficients, $T_a$ and $T_b$, the covariance matrix of the initial state, and the coherent amplitudes.
	For correlated transmission coefficients, $\langle\Delta T_a\Delta T_b\rangle\neq 0$, this term may attain negative values.
	
\subsection{Higher-order non-Gaussian test}
	Let us analyze a higher-order moments test for the turbulent atmosphere.
	In particular, we consider an entanglement condition which includes moments up to the fourth order,
	\begin{align}\label{eq:HO}
		S=\begin{vmatrix}
			1 & \langle\hat a\hat b^\dagger\rangle\\
			\langle\hat a^\dagger\hat b\rangle & \langle\hat a^{\dagger}\hat a\hat b^{\dagger}\hat b\rangle.
		\end{vmatrix}\stackrel{\text{ent.}}{<}0,
	\end{align}
	which is based on a $2\times 2$ minor of the partial transposed matrix of moments~\eqref{eq:MPT}.
	In Ref.~\cite{PNAS09}, this particular condition was used to experimentally verify non-Gaussian entanglement of a state which is invisible to all second-order (Gaussian) entanglement tests.
	From the partially transposed matrix of moments at the receiver, cf. Eq.~\eqref{eq:MoutPT}, we see that the output condition of the test~\eqref{eq:HO} can be given in the form
	\begin{align}\label{eq:HOout}
		S^{\rm out}=\langle T_a^2T_b^2\rangle\left[ S+\Gamma^{(1)}_{\{a,b\}}|\langle \hat a\hat b^\dagger\rangle|^2\right],
	\end{align}
	with $\Gamma_{\{a,b\}}^{(1)}$ defined in Eq.~\eqref{eq:Gammamulti}.
	Again, we observe that the turbulence adds a positive term to a possibly negative $S$, which eventually may lead to the fact that the non-Gaussian entanglement cannot be verified by this test anymore.

	Now, we apply this higher-order entanglement condition~\eqref{eq:HOout} to the specific state
	\begin{align}\label{eq:ECS}
		|\Psi^-\rangle=\mathcal{N}(\alpha,\beta)(|\alpha\rangle\otimes|\beta\rangle-|-\alpha\rangle\otimes|-\beta\rangle)	
	\end{align}
	with $\mathcal{N}(\alpha,\beta)=[2(1-e^{-2(|\alpha|^2+|\beta|^2)})]^{-1/2}$.
	This kind of non-Gaussian state belongs to the family of so-called entangled coherent states~\cite{ECS} which have been experimentally realized; see, e.g., Ref.~\cite{ECSexperiment}.
	The state satisfies the entanglement condition~\eqref{eq:HO}, as $S=-16|\alpha|^2|\beta|^2\mathcal{N}(\alpha,\beta)^4e^{-2(|\alpha|^2+|\beta|^2)}$, which is negative for all $\alpha\neq 0\neq \beta$.
	Hence, the entanglement of the state can only be detected by higher-order criteria, which means it exhibits a form of genuinely non-Gaussian entanglement~\cite{ShchukinVogel2005}.
	To examine the state after passing a turbulent loss medium, the corresponding test can be written as $S^{\rm out}<0$, with
	\begin{align}\label{eq:SoutECS}
		S^{\rm out}=\langle T_a^2T_b^2\rangle S\left[-1+\Gamma^{(1)}_{\{a,b\}}\frac{(1+e^{-2(|\alpha|^2+|\beta|^2)})^2}{4e^{-2(|\alpha|^2+|\beta|^2)}}\right].
	\end{align}
	Again, the perturbing effect on the entanglement depends on the fluctuation parameter $\Gamma^{(1)}_{\{a,b\}}$.

\begin{figure}[ht]
		\includegraphics[width=1\linewidth]{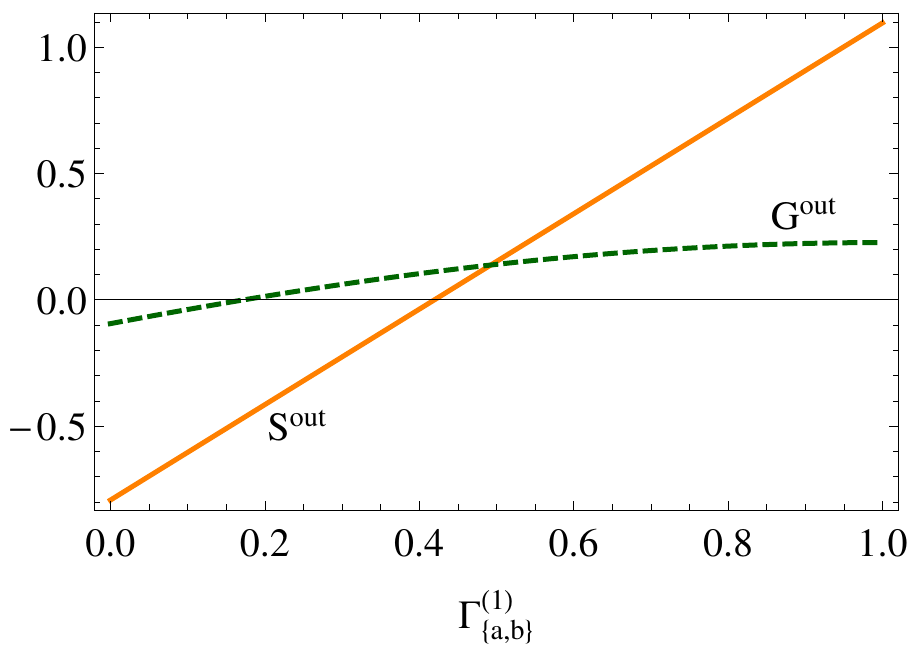}
		\caption{(Color online)
		The non-Gaussian entanglement test $S^{\rm out}<0$ [cf. Eq.~\eqref{eq:HOout}] for the entangled coherent states~\eqref{eq:ECS} with $\alpha=\beta$ (solid line) and the Gaussian entanglement condition $G^{\rm out}<0$ [cf. Eq.~\eqref{eq:IOR_Simon}] of a $5.7$dB squeezed two-mode squeezed-vacuum state (dashed line) are plotted as a function of the fluctuation parameter $\Gamma^{(1)}_{\{a,b\}}$.
		The fluctuating loss is considered as uncorrelated, with the parameters 
		$\langle T^2_a\rangle{=}\langle T^2_b\rangle{=}0.7$ and $\langle T_a\rangle{=}\langle T_b\rangle{=}\langle T^2_a\rangle\sqrt{1-\Gamma^{(1)}_{\{a,b\}}}$.
		Both states have the same mean photon number.
		}\label{fig:nonGauss}
	\end{figure}

	An interesting question is how robust non-Gaussian entanglement is against fluctuating losses compared to Gaussian one.
	In Fig.~\ref{fig:nonGauss}, the entanglement of a Gaussian two-mode squeezed-vacuum state ($5.7\,$dB squeezing) is compared with the non-Gaussian state~\eqref{eq:ECS} with $\alpha=\beta$.
	In order to assure the comparability, both states are chosen to have the same total photon number 
	$\langle\hat n_a+\hat n_b\rangle{=}1$, as we already discovered that the photon number is crucial for 
	the transfer of nonclassical states through fluctuating-loss channels.
	To test for the entanglement of the two-mode squeezed-vacuum state, we use the output version of the Simon criterion as derived in~\cite{BSSV16}, cf. also Eq.~(\ref{eq:IOR_Simon}).
	From Fig.~\ref{fig:nonGauss}, we conclude that the non-Gaussian entanglement is preserved for a larger range of fluctuating loss, quantified by $\Gamma^{(1)}_{\{a,b\}}$, compared to the entanglement of the Gaussian state.
	Thus, higher-order entanglement might be more robust under fluctuating losses than Gaussian entanglement.
	Therefore, quantum communication or teleportation that employ non-Gaussian correlations might be favorable for implementations in free-space channels.

\subsection{Multimode entanglement}
	After studying bipartite non-Gaussian entanglement in turbulent atmospheric channels, now we also consider multimode entanglement under such conditions.
	Therefore, and as a proof of principle, we analyze a four-mode scenario, where each mode is represented by the operator $\hat a_i$ with $i=1,2,3,4$.
	As entanglement tests we will use the partial transpositions of the multimode matrix of moments~\eqref{eq:MPT}.
	In particular, we apply the partial transpositions to the four-mode $2\times2$ minors $m_{(1,2;3,4)}$, $m_{(1,3;2,4)}$, and $m_{(2,3;1,4)}$, where $m_{(i,j;k,l)}$ is given by
	\begin{align}
		\label{eq:m}
		m_{(i,j;k,l)}=
			\begin{vmatrix} 
			\langle\hat a_i^\dagger\hat a_i\hat a_j^\dagger\hat a_j\rangle & \langle\hat a_i^\dagger\hat a_j^\dagger\hat a_k\hat a_l\rangle\\
			\langle\hat a_i\hat a_j\hat a_k^\dagger\hat a_l^\dagger\rangle & \langle\hat a_k^\dagger\hat a_k\hat a_l^\dagger\hat a_l\rangle
			\end{vmatrix}.
	\end{align}
	As the $N$ modes can be separated in $2^{N-1}-1$ different nontrivial bipartitions, we have to consider in our case seven different entanglement conditions.
	These conditions are given by the negativity of $m_{(1,2;3,4)}^{\{1\}}$, $m_{(1,2;3,4)}^{\{2\}}$, $m_{(1,2;3,4)}^{\{3\}}$, $m_{(1,2;3,4)}^{\{1,2,3\}}$, $m_{(1,2;3,4)}^{\{1,2\}}$, $m_{(1,3;2,4)}^{\{1,3\}}$, and $m_{(2,3;1,4)}^{\{2,3\}}$, where the superscript denotes in which partition the transposition is carried out.
	Now we will introduce the corresponding conditions after passing through a fluctuating-loss channel.
	Hence, we first formulate the output form of $m_{(i,j;k,l)}$,
	\begin{align}
	\begin{aligned}
		\label{eq:mout}
		m_{(i,j;k,l)}^{\rm out}=&\langle T_i^2T_j^2\rangle \langle T_k^2T_l^2\rangle\\
		&\times [m_{(i,j;k,l)}+\Gamma_{\{i,j;k,l\}}^{( 1)}|\langle\hat a_i^\dagger\hat a_j^\dagger\hat a_k\hat a_l\rangle|^2],
	\end{aligned}
	\end{align}
	with $\Gamma_{\{i,j;k,l\}}^{( 1)}$ defined in Eq.~\eqref{eq:Gammamulti}.

	As demonstrated in our previous analysis, here we also observe that the turbulence adds a positive term to the entanglement test.
	Due to the turbulence, one may not be able to prove entanglement with this test anymore.
	It is worth mentioning that, for the structure of $m_{(i,j;k,l)}^{\rm out}$, it is not important in which modes the fluctuating losses occur, it is just important that there is a fluctuating loss in at least one mode.
	However, for the special case that all modes undergo a correlated loss, that is 
	$T{=}T_1{=}T_2{=}T_3{=}T_4$ and $\Gamma_{\{i,j;k,l\}}^{( 1)}=0$, we see that the output is simply scaled as $m_{(i,j;k,l)}^{\rm out}=\langle T^4\rangle^2 m_{(i,j;k,l)}$.
	This means that correlated loss cannot affect the sign of the partial transposition of $m_{(i,j;k,l)}^{\rm out}$ and, hence, entanglement will be preserved.
	We already observed a similar behavior for two-mode nonclassical correlations; cf. Sec.~\ref{ch:multi-mode-nc}.
	In the present scenario, one may also artificially correlate the modes via the protocol introduced in Ref.~\cite{BSSV16}, in order to preserve the non-Gaussian entanglement.

	Let us now study the behavior of the test including the contributions of the atmosphere [cf. Eq.~\eqref{eq:mout}] for a particular state and under realistic conditions.
	We consider a state of the form 
	\begin{align}\label{eq:contW}
		|\psi(\alpha)\rangle={\mathcal N(\alpha)}\sum_{i=1}^N
		|\alpha\rangle^{\otimes(i-1)}\otimes|-\alpha\rangle\otimes|\alpha\rangle^{\otimes(N-i)},
	\end{align}
	with $N=4$ in our case, a proper normalization constant $\mathcal N(\alpha)$, and the definition $|\phi\rangle^{\otimes 0}\equiv 1$.
	Such a state is a continuous variable version of the $W$ state.
	A way of implementing such a state was recently proposed in Ref.~\cite{Wstate}.
	For this particular state, the seven different partially transposed minors reduce to only the two following forms~\cite{SV06RC}:
	$m_I{=}m_{(1,2;3,4)}^{\{1\}}{=}m_{(1,2;3,4)}^{\{2\}}{=}m_{(1,2;3,4)}^{\{3\}}{=}m_{(1,2;3,4)}^{\{1,2,3\}}$ and $m_{II}{=}m_{(1,2;3,4)}^{\{1,2\}}{=}m_{(1,3;2,4)}^{\{1,3\}}{=}m_{(2,3;1,4)}^{\{2,3\}}$.
	The same holds for $m_{(i,j;k,l)}^{{\rm out}}$.
	While for $m_I$ the entanglement between one mode and the rest of the modes is considered, $m_{II}$ corresponds to an equal-sized partitioning.

\begin{figure}[ht]
		\includegraphics[width=1\linewidth]{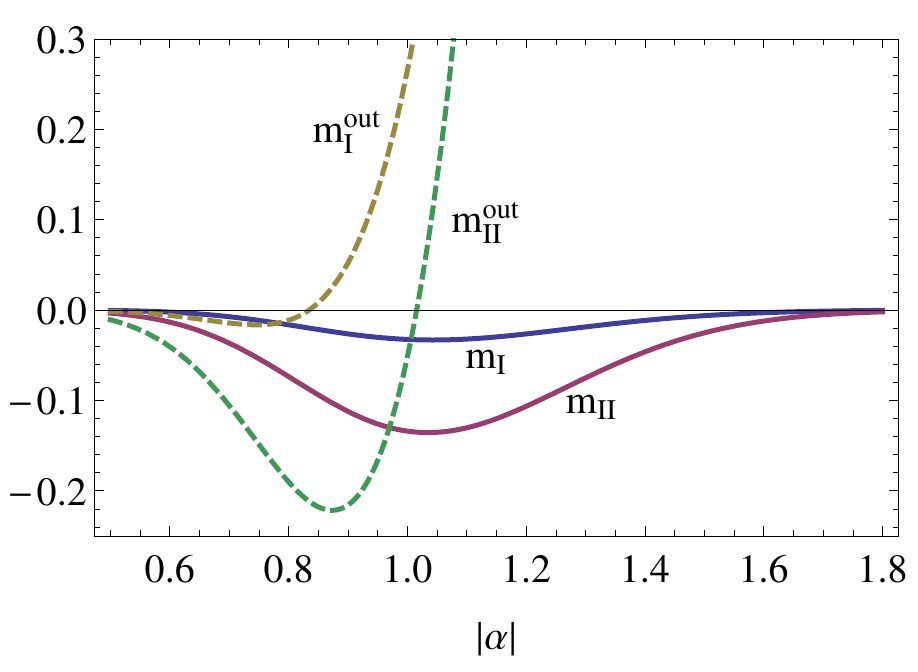}
		\caption{(Color online)
			The four-mode entanglement input tests $m_{I}<0$ and $m_{II}<0$ and their output versions $m_{I}^{{\rm out}}<0$ and $m_{II}^{{\rm out}}<0$ for the state~\eqref{eq:contW} are displayed as a function of the coherent amplitude $|\alpha|$.
			$m_{I}^{{\rm out}}$ and $m_{II}^{{\rm out}}$ are scaled by a factor of $5\times10^3$.
			The fluctuating losses are uncorrelated, with $\langle T_i\rangle=0.398$, $\langle T_i^2\rangle=0.163$ for $i=1,2,3,4$ and $\Gamma_{\{i,j;k,l\}}^{(1)}=0.119$.
		}\label{fig:multimode}
	\end{figure}

	In Fig.~\ref{fig:multimode}, the entanglement tests for the undisturbed input and the attenuated output test are plotted as a function of the coherent amplitude $|\alpha|$ of the state~\eqref{eq:contW}.
	We consider equal but uncorrelated losses in each mode.
	As in Sec.~\ref{ch:multi-mode-nc}, we apply the beam-wandering model~\cite{beamwandering,VSV2016}.
	The first conclusion that can be obtained from Fig.~\ref{fig:multimode} is that turbulence reduces the range of the coherent amplitude for which we can still detect entanglement.
	This is consistent with the results obtained above for other quantum effects---that is, states with higher mean photon numbers are more fragile to fluctuating losses.
	Second, we observe that $m_{II}^{{\rm out}}$ is more robust against the fluctuating losses than $m_{I}^{{\rm out}}$.
	Hence, the entanglement between a symmetric splitting of the modes, probed by $m_{II}^{{\rm out}}$, is more robust then the entanglement between one mode and the other three, which are tested by $m_{I}^{{\rm out}}$.
	Therefore, our analysis can determine which kind of entanglement is more stable and, hence, preferable in atmospheric links.
	The possibility to assess multimode quantum correlations under such fluctuating losses is an important tool to design and develop free-space quantum networks.

\section{Balanced homodyne correlation measurement for atmospheric channels}\label{ch:measurement}
	In the previous parts of this article, we focused on nonclassicality and entanglement criteria based on the matrix of moments for quantum states that are subjected to attenuations in atmospheric fluctuating-loss channels.
	Such criteria are only of practical interest if one has measurement principles at hand to determine the corresponding moments.
	In this section, we briefly outline a measurement strategy to experimentally access the desired quantities.
	In particular, this approach will be based on the method of homodyne correlation measurements~\cite{Vogel91,Vogel95,SV06} and it will make use of a local oscillator (LO) field which co-propagates with the signal field through the loss channel~\cite{Elser,Semenov2012}.
	For simplicity, we will focus on a single-mode detection scheme, which, however, can be straightforwardly extended to the multimode cases~\cite{SV06}.

	To ensure a proper performance of the measurement, a good interference of LO and signal is crucial.
	Therefore, the LO is propagating with the signal field in the same spatial mode but in an orthogonal polarization mode.
	As the depolarization effects of the atmosphere are negligibly small~\cite{Tatarskii}, both fields, LO and signal, experience the same atmospheric disturbances.
	Hence, the atmospheric attenuation of the spatial mode profiles are equal for both modes, which leads to an optimal interference of LO and signal field.
	Furthermore, the LO acts as a spatial and spectral filter, which even allows for daylight operation of the proposed scheme.
	Note that by measuring the intensity of the LO, one can directly monitor the channel loss.
	This enables one to apply post-selection~\cite{beamwandering} or to correlate two or more channel transmissions~\cite{BSSV16}.

\begin{figure}[ht]
		\includegraphics[width=0.95\linewidth]{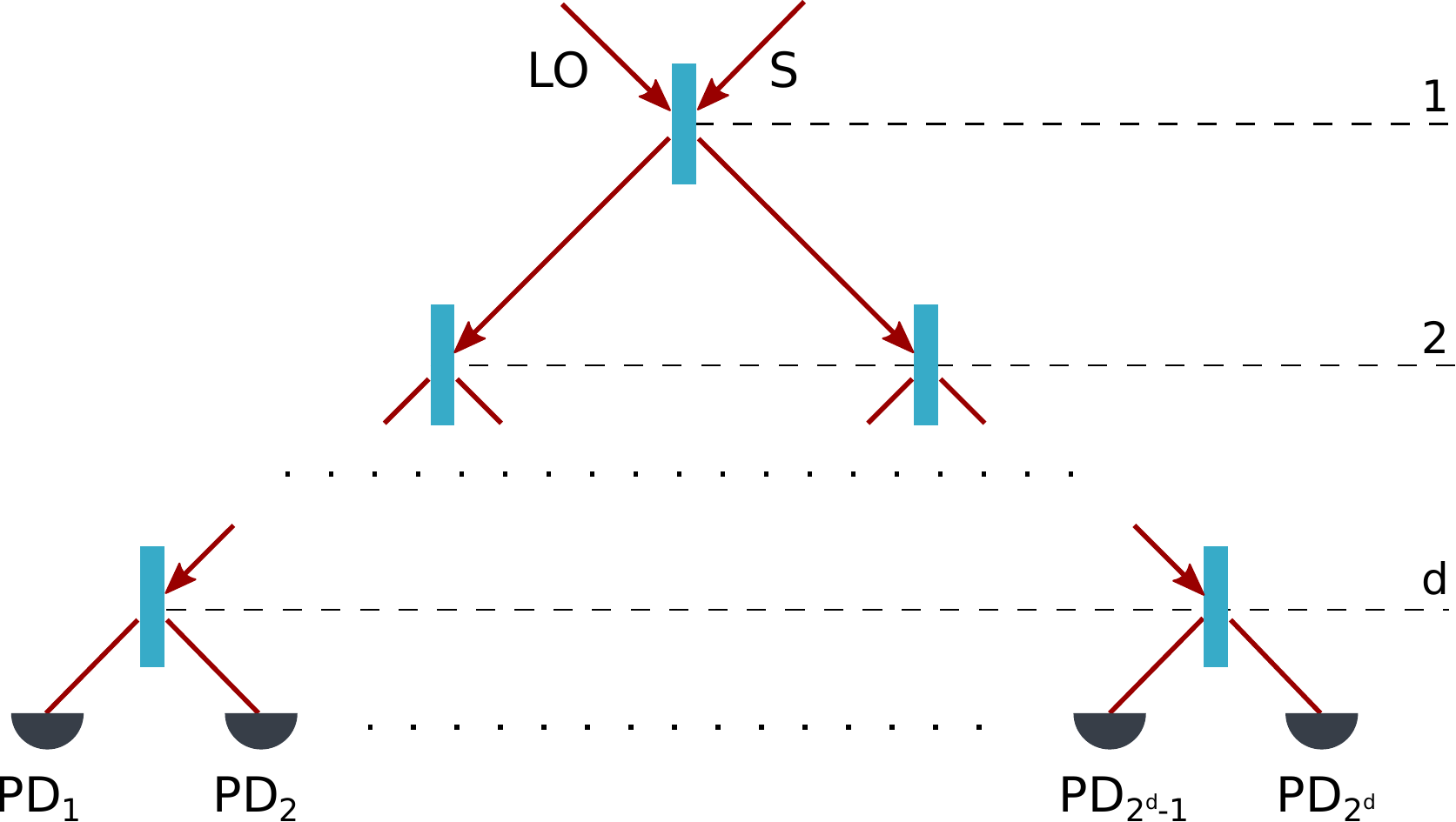}
		\caption{(Color online)
			The setup for a homodyne correlation measurement is shown.
			All beam splitters are symmetric and the signal is detected by $2^d$ photodetectors, where $d$ is the depth of the setup.
		}\label{fig:MD}
	\end{figure}

	After we briefly discussed how the signal and LO can be propagated through the atmosphere, we now focus on the actual measurement technique.
	In particular, we will show how the moments $\langle \hat a^{\dagger n} \hat a^m\rangle_{\rm out}$ of the signal field including channel loss can be measured with the device shown in Fig.~\ref{fig:MD}.
	The setup consists of $50:50$ beam splitters and photodetectors.
	At the first beam splitter, the signal is superimposed with the LO, which is prepared in a coherent state $|\alpha\rangle$ with $\alpha=|\alpha|e^{i\varphi_{\rm LO}}$.
	Subsequently, the outputs of this first beam splitter are further equally split.
	The resulting beams are measured with $2^d$ photodetectors, where $d$ is depth of the measurement device; cf.~Fig.~\ref{fig:MD}.
	This allows us to measure moments with $n+m\leq 2^{d-1}$~\cite{SV06}.

	As an example, we focus on the case $d=2$ for which we have four different detection modes which are described by the operators $\hat a_1, \hat a_2, \hat a_3$, and $\hat a_4$.
	In this scenario, higher-order correlations can be detected, for example, by measuring
	\begin{align}
	\begin{aligned}
		&\langle \hat a_1^\dagger\hat a_1 \hat a_2^\dagger\hat a_2\rangle_{\rm out}-2\langle \hat a_1^\dagger\hat a_1\hat a_3^\dagger\hat a_3\rangle_{\rm out}+\langle  \hat a_3^\dagger\hat a_3 \hat a_4^\dagger\hat a_4\rangle_{\rm out}=\\
		&\frac{1}{4}|\alpha|^2\big(\langle \hat a^2\rangle_{\rm out} {e}^{-2{i} \varphi}+2\langle\hat a^\dagger \hat a\rangle_{\rm out}+\langle \hat a^{\dagger 2}\rangle_{\rm out} { e}^{2{ i} \varphi}\big),
	\end{aligned}
	\end{align}
	with $\varphi=\varphi_{\rm LO}+\pi/2$.
	By a Fourier transformation with respect to $\varphi$, one can retrieve moments such as $\langle \hat a^2\rangle_{\rm out}$.
	In this way, we directly obtain the output moments after the transmission through the atmosphere, which constitute the output matrix of moments.
	In a similar way, all moments $\langle \hat a^{\dagger n} \hat a^m\rangle_{\rm out}$ can be obtained by such a balanced homodyne correlation measurement.
	As already mentioned, the extension to multimode moments and more details on this treatment can be found in Ref.~\cite{SV06}.
	Therefore, the layout in Fig.~\ref{fig:MD} can be used to measure the moments of single- and multi-mode matrices of moments.

\section{Summary}\label{ch:Summary}
	In summary, we have studied moment-based nonclassicality and entanglement criteria in the presence of fluctuating losses.
	We introduced general input-output relations for the single- and multi-mode matrix of moments.
	In contrast to deterministic losses, the output matrices show a nontrivial dependency on fluctuation parameters.
	Particularly, different orders of moments scale with different moments of the transmission coefficients.
	Additionally, the turbulence can lead to a mixing of different moments.
	Both effects may diminish the quantumness of the radiation field.
	We also introduced correlation parameters, which quantify the strength of fluctuating losses.
	
	Based on general input-output relations, we studied the corresponding nonclassicality conditions in terms of normally ordered moments.
	For the single-mode case, we could identify that the mean photon number of the quantum state in atmospheric channels can be a crucial parameter.
	More precisely, states with lower mean photon numbers turn out to be more robust against fluctuating losses.
	We could observe the same behavior also for multimode nonclassicality and entanglement, which indicates the general character of this effect.
	Another effect is the dependency of the nonclassicality on the coherent displacement of the considered quantum state.
	We have shown that, for a displaced squeezed state, the nonclassicality transfer in free-space links strictly depends on the direction of the displacement.
	This allows one to optimize the state in such scenarios.
	In the case of multimode nonclassicality, we could show that correlated attenuations can preserve nonclassical correlations.
	Thus, artificially correlating the losses of different modes, as proposed in Ref.~\cite{BSSV16}, can also lead to a preservation of general quantum correlations.

	Besides the nonclassical properties of radiation fields, we also examined entanglement conditions based on the negativity of the partial transposition of the output state of turbulent loss channels.
	We studied nontrivial scenarios of non-Gaussian and multimode entanglement.
	We were able to show that bipartite non-Gaussian entanglement can be more robust against fluctuating losses than Gaussian entanglement.
	Therefore, quantum communication strategies in free-space based on non-Gaussian entanglement might be advantageous compared with others based on Gaussian entanglement.
	Furthermore, we demonstrated the feasibility of verifying multimode entanglement with criteria based on the matrix of moments at the receivers.
	Finally, we proposed how one can actually measure the moments needed for the various criteria.
	For this purpose, we adapted the method of balanced homodyne correlation measurement to the atmospheric scenarios.

\subsection*{Acknowledgements}
	The authors gratefully acknowledge financial support by the Deutsche Forschungsgemeinschaft through Project No. VO 501/22-1.


\begin{thebibliography}{99}
\bibitem{Ursin}
	R.~Ursin \textit{et al.}, 
	Entanglement-based quantum communication over 144 km,
	Nat. Phys. \textbf{3}, 481 (2007).
\bibitem{Scheidl} 
	T.~Scheidl \textit{et al.}, 
	Feasibility of 300km quantum key distribution with entangled states, 
	New J. Phys. \textbf{11}, 085002 (2009).
\bibitem{Fedrizzi2009}
	A.~Fedrizzi, R.~Ursin, T.~Herbst, M.~Nespoli, R.~Prevedel, T.~Scheidl, F.~Tiefenbacher, T.~Jennewein, and A.~Zeilinger,
	High-fidelity transmission of entanglement over a high-loss free-space channel,
	Nat. Phys. {\bf 5}, 389 (2009).
\bibitem{Capraro}
	I.~Capraro, A.~Tomaello, A.~Dall'Arche, F.~Gerlin, R.~Ursin, G.~Vallone, and P.~Villoresi, 
	Impact of Turbulence in Long Range Quantum and Classical Communications, 
	Phys. Rev. Lett. \textbf{109}, 200502 (2012).
\bibitem{Yin} 
	J.~Yin \textit{et al.}, 
	Quantum teleportation and entanglement distribution over 100-kilometre free-space channels, 
	Nature (London) \textbf{488}, 185 (2012).
\bibitem{Ma} 
	X.~Ma \textit{et al.},
	Quantum teleportation over 143 kilometres using active feed-forward,
	Nature (London) \textbf{489}, 269 (2012).
\bibitem{Peuntinger}  
	C.~Peuntinger, B.~Heim, Ch.~M\"uller, Ch.~Gabriel, Ch.~Marquardt, and G.~Leuchs, 
	Distribution of Squeezed States through an Atmospheric Channel, 
	Phys. Rev. Lett. \textbf{113}, 060502 (2014).
\bibitem{Villoresi08}
	P. Villoresi, T. Jennewein, F. Tamburini, M. Aspelmeyer, C. Bonato, R. Ursin, C. Pernechele, V. Luceri, G. Bianco, and A. Zeilinger,
	Experimental verification of the feasibility of a quantum channel between space and Earth,
	New J. Phys. \textbf{10}, 033038 (2008).
\bibitem{Vallone15} 
	G. Vallone, D. Bacco, D. Dequal, S. Gaiarin, V. Luceri, G. Bianco, and P. Villoresi,
	Experimental Satellite Quantum Communications,
	Phys. Rev. Lett. {\bf 115}, 040502 (2015).
\bibitem{Dequal16}
	D. Dequal, G. Vallone, D. Bacco, S. Gaiarin, V. Luceri, G. Bianco, and P. Villoresi, 
	Experimental single photon exchange along a space link of 7000 km,
	Phys. Rev. A {\bf 93}, 010301(R) (2016).
\bibitem{Vallone16}
	G. Vallone, D. Dequal, M. Tomasin, F. Vedovato, M. Schiavon, V. Luceri, G. Bianco, and P. Villoresi, 
	Interference at the Single Photon Level Along Satellite-Ground Channels,
	Phys. Rev. Lett. {\bf 116}, 253601 (2016).
\bibitem{Semenov2009} 
	A. A. Semenov and W. Vogel, 
	Quantum light in the turbulent atmosphere,
	Phys. Rev.  A {\bf 80}, 021802(R) (2009).
\bibitem{beamwandering} 
	D. Yu. Vasylyev, A. A. Semenov, and W. Vogel, 
	Toward Global Quantum Communication: Beam Wandering Preserves Nonclassicality,
	Phys. Rev. Lett. \textbf{108}, 220501 (2012).
\bibitem{VSV2016} 
	D. Yu. Vasylyev, A. A. Semenov, and W. Vogel, 
	Atmospheric Quantum Channels with Weak and Strong Turbulence,
	Phys. Rev. Lett. \textbf{117}, 090501 (2016).
\bibitem{Usenko} 
	V. C. Usenko, B. Heim, C. Peuntinger, C. Wittmann, C. Marquardt, G. Leuchs, and R. Filip, 
	Entanglement of Gaussian states and the applicability to quantum key distribution over fading channels,
	New J. Phys. \textbf{14}, 093048 (2012).
\bibitem{Elser} 
	D. Elser, T. Bartley, B. Heim, C. Wittmann, D. Sych, and G. Leuchs,
	Feasibility of free space quantum key distribution with coherent polarization states,
	New J. Phys. \textbf{11}, 045014 (2009).
\bibitem{Semenov2012} 
	 A. A. Semenov, F. T\"{o}ppel, D. Yu. Vasylyev, H. V. Gomonay, and W. Vogel,
	Homodyne detection for atmosphere channels,
	Phys. Rev. A \textbf{85}, 013826 (2012).
\bibitem{Croal16} 
	C. Croal, C. Peuntinger, B. Heim, I. Khan, C. Marquardt, G. Leuchs, P. Wallden, E. Andersson, and N. Korolkova,
	Free-Space Quantum Signatures Using Heterodyne Measurements,
	Phys. Rev. Lett. \textbf{117}, 100503 (2016).
\bibitem{Semenov2010} 
	A. A. Semenov and W. Vogel, 
	Entanglement transfer through the turbulent atmosphere,
	Phys. Rev. A {\bf 81},  023835 (2010).
\bibitem{GaussSatellites}
	N. Hosseinidehaj and R. Malaney,
	Gaussian entanglement distribution via satellites,
	Phys. Rev. A \textbf{91}, 022304 (2015).
\bibitem{Bohmann15}
	M. Bohmann, J. Sperling, and W. Vogel,
	Entanglement and phase properties of noisy N00N states,
	Phys. Rev. A \textbf{91}, 042332 (2015). 
\bibitem{BSSV16}
	M. Bohmann, A. A. Semenov, J. Sperling, and W. Vogel,
	Gaussian entanglement in the turbulent atmosphere,
	Phys. Rev. A {\bf 94}, 010302(R) (2016).
\bibitem{Simon2000}
	R. Simon,
	Peres-Horodecki Separability Criterion for Continuous Variable Systems,
	Phys. Rev. Lett. \textbf{84}, 2726 (2000).
\bibitem{AgTa} 
	G. S. Agarwal and K. Tara, 
	Nonclassical character of states exhibiting no squeezing or sub-Poissonian statistics, 
	Phys. Rev. A {\bf 46} 485 (1992).
\bibitem{Majorization}
	I. I. Arkhipov, J. Pe\v{r}ina, O. Haderka, and V.Mich\'alek,
	Experimental detection of nonclassicality of single-mode fields via intensity moments,
	Opt. Express {\bf 24}, 29496 (2016).
\bibitem{Ag} 
	G. S. Agarwal, 
	Nonclassical characteristics of the marginals for the radiation field,
	Opt. Commun. {\bf 95}, 109 (1993).
\bibitem{SV05nca}
	E. Shchukin, Th. Richter, and W. Vogel,
	Nonclassicality criteria in terms of moments,
	Phys. Rev. A {\bf 71}, 011802(R) (2005)
\bibitem{SV05ncb}
	E. Shchukin and W. Vogel,
	Nonclassicality moments and their measurements,
	Phys. Rev. A {\bf 72}, 043808 (2005).
\bibitem{Miran} 
	A. Miranowicz, M. Bartkowiak, X. Wang, Y.-x. Liu, and F. Nori, 
	Testing nonclassicality in multimode fields: A unified derivation of classical inequalities,
	Phys. Rev. A {\bf 82}, 013824 (2010).
\bibitem{ShchukinVogel2005}
	E. Shchukin and W. Vogel,
	Inseparability Criteria for Continuous Bipartite Quantum States,
	Phys. Rev. Lett. \textbf{95}, 230502 (2005).
\bibitem{SV06RC}
	E. Shchukin and W. Vogel,
	Conditions for multipartite continuous-variable entanglement,
	Phys. Rev. A {\bf 74}, 030302(R) (2006).
\bibitem{Sudarshan}
	E. C. G. Sudarshan,
	Equivalence of Semiclassical and Quantum Mechanical Descriptions of Statistical Light Beams, 
	Phys. Rev. Lett. {\bf 10}, 277 (1963).
\bibitem{Glauber} 
	R. J. Glauber, 
	Coherent and incoherent states of the radiation field, 
	Phys. Rev. {\bf 131}, 2766 (1963).
\bibitem{Short} 
	R. Short and L. Mandel, 
	Observation of Sub-Poissonian Photon Statistics, 
	Phys. Rev. Lett. {\bf 51}, 384 (1983).	
\bibitem{MaQ}
	L. Mandel, 
	Sub-Poissonian photon statistics in resonance fluorescence,
	Opt. Lett. {\bf 4}, 205 (1979).
\bibitem{Hillery87}
	M. Hillery,
	Amplitude-squared squeezing of the electromagnetic field,
	Phys. Rev. A {\bf 36} 3796 (1987). 
\bibitem{SV05HOAS}
	E. Shchukin and W. Vogel,
	Higher-order amplitude squeezing,
	J. Phys: Conference Series {\bf 36}, 183 (2006).  
\bibitem{Buch}
	Werner Vogel and Dirk-Gunnar Welsch,
	{\it Quantum optics}
	(Wiley-VCH, Berlin, 2006).
\bibitem{Agarwal13}
	G. S. Agarwal,
	{\it Quantum Optics},
	(Cambridge University, Cambridge, 2013).
\bibitem{Agudelo13}
	E. Agudelo, J. Sperling, and W. Vogel,
	Quasiprobabilities for multipartite quantum correlations of light,
	Phys. Rev. A {\bf 87}, 033811 (2013).
\bibitem{Peres}
	A. Peres, 
	Separability Criterion for Density Matrices,
	Phys. Rev. Lett. {\bf 77}, 1413 (1996).
\bibitem{Horodecki96}
	M. Horodecki, P. Horodecki, and R. Horodecki, 
	Separability of mixed states: necessary and sufficient conditions,
	Phys. Lett. A {\bf 223}, 1 (1996).
\bibitem{Gerke16}
	S. Gerke, J. Sperling, W. Vogel, Y. Cai, J. Roslund, N. Treps, and C. Fabre,
	Multipartite Entanglement of a Two-Separable State,
	Phys. Rev. Lett. {\bf 117}, 110502 (2016).
\bibitem{SV13}
	J. Sperling and W. Vogel,
	Multipartite Entanglement Witnesses,
	Phys. Rev. Lett. {\bf 111}, 110503 (2013).
\bibitem{Gerke15}
	S. Gerke, J. Sperling, W. Vogel, Y. Cai, J. Roslund, N. Treps, and C. Fabre,
	Full Multipartite Entanglement of Frequency-Comb Gaussian States,
	Phys. Rev. Lett. {\bf 114}, 050501 (2015).
\bibitem{Barbosa}
	F. A. S. Barbosa, A. J. de Faria, A. S. Coelho, K. N. Cassemiro, A. S. Villar, P. Nussenzveig, and M. Martinelli,
	Disentanglement in bipartite continuous-variable systems,
	Phys. Rev. A {\bf 84}, 052330 (2011).
\bibitem{PNAS09}
	R. M. Gomes, A. Salles, F. Toscano, P. H. Souto Ribeiro, and S. P. Walborn,
	Quantum entanglement beyond Gaussian criteria,
	PNAS {\bf 106} 21517 (2009).
\bibitem{ECS}
	B. C. Sanders,
	Review of entangled coherent states,
	J. Phys. A: Math. Theor. 45, 244002 (2012).
\bibitem{ECSexperiment}
	A. Ourjoumtsev, F. Ferreyrol, R. Tualle-Brouri, and P. Grangier,
	Preparation of non-local superpositions of quasi-classical light states,
	Nat. Phys. {\bf 5}, 189 (2009). 
\bibitem{Wstate}
	T. Liu, Q.-P. Su, S.-J. Xiong, J.-M. Liu, C.-P. Yang, and F. Nori,
	Generation of a macroscopic entangled coherent state using quantum memories in circuit QED,
	Sci. Rep. {\bf 6}, 32004 (2016). 
\bibitem{Vogel91}
	W. Vogel, 
	Squeezing and Anomalous Moments in Resonance Fluorescence,
	Phys. Rev. Lett. {\bf 67}, 2450 (1991).
\bibitem{Vogel95}
	W. Vogel,
	Homodyne correlation measurements with weak local oscillators,
	Phys. Rev. A {\bf 51}, 4160 (1995). 
\bibitem{SV06}
	E. Shchukin and W. Vogel,
	Universal Measurement of Quantum Correlations of Radiation,
	Phys. Rev. Lett. {\bf 96}, 200403 (2006).
 \bibitem{Tatarskii} 
	V. Tatarskii, 
	{\it Effects of the Turbulent Atmosphere on Wave Propagation},
	(IPST, Jerusalem, 1972).

\end{thebibliography}
\end{document}